\documentclass[12pt]{iopart}

\usepackage{graphicx}
\usepackage{lineno}
\usepackage{hyperref}
\hypersetup{colorlinks,linkcolor={blue},citecolor={blue},urlcolor={blue}}
\usepackage{tcolorbox}
\usepackage{subfigure}
\begin{document}


\title[]{Electrospray Thruster Plume Dynamics: Insights from Precise PP Coulomb Field Simulation}

\author[]{Zhe Liu, Yinjian Zhao*}
\ead{zhaoyinjian@hit.edu.cn}
\address{School of Energy Science and Engineering, Harbin Institute of Technology, Harbin 150001, People’s Republic of China}

\vspace{10pt}
\begin{indented}
\item[]\today
\end{indented}

\begin{abstract}
Electrospray thrusters are one important type of micropropulsion systems being developed for next-generation space missions, yet the primary challenge to their operational lifespan is propellant overspray resulting from wide plume angles driven by Coulomb interactions among charged droplets. While existing models often employ truncated Coulomb field approximations, such simplifications compromise accuracy in predicting divergence dynamics. In this study, a particle-particle (PP) simulation method is used to directly calculate the interactions between droplets in an electrospray plume coupled with background electric field effects for simulation. The model integrates Boris pusher for numerical integration, validated through binary collision tests verification. Parametric analysis systematically evaluates six key variables—droplet charge, droplet mass, emission interval, droplet initial velocity, and electric field components—to quantify their impacts on plume divergence.
The shape of the simulated electrospray plume and the velocity of the droplets in it are analyzed. Parametric analysis demonstrate that reducing droplet charge, increasing droplet mass, extending emission time intervals, and elevating initial drift velocity collectively reduce plume half-angle. These results quantitatively establish parameter-plume relationships, providing direct guidance for thruster optimization.
\end{abstract}

\section{Introduction} \label{sec:intro}

With the rapid development of micro-nano satellites and high-precision space exploration missions, electric propulsion systems have been established as the core technology for spacecraft attitude and orbit control \cite{gomezjenkinsCubeSatConstellationManagement2018, oreillyElectricPropulsionMethods2021, mazouffreElectricPropulsionSatellites2016}, owing to their high specific impulse (typically ranging from 1000 s to 3500 s \cite{goebel2023fundamentals}, whereas traditional chemical propulsion systems rarely exceed 500 s \cite{lozanoDesignMicrostructuringMaterials2024}) and superior efficiency. Among these systems, electrospray thrusters demonstrate unique advantages in ultra-precision space missions such as gravitational wave detection \cite{wirz2019electrospray}, attributed to their micro-Newton-level thrust accuracy and ultrahigh specific impulse (up to 4000 s). However, thruster longevity, a critical engineering parameter, is significantly compromised by electrode sputtering contamination and reverse current effects \cite{thuppulLifetimeConsiderationsElectrospray2020}, both induced by plume expansion during operation. It is of significant engineering importance to elucidate the dynamic mechanisms underlying electrospray plume expansion, as this understanding is pivotal for extending thruster service cycles and optimizing the reliability of space propulsion systems.

The evolution of electrospray plumes is governed by the coupling of multiple parameters, including droplet charge, mass, emission time intervals, and initial emission velocities \cite{millerCapillaryIonicLiquid2021}. It has been demonstrated that Coulomb repulsion between droplets serves as the dominant factor driving plume expansion \cite{breddanElectrosprayPlumeEvolution2023, grifollEfficientLagrangianSimulation2012} . Nevertheless, experimental approaches are inherently constrained by material limitations and systemic integration effects, rendering independent modulation and quantitative characterization of single variables unfeasible. Consequently, numerical simulation has emerged as a vital complementary tool for mechanistic investigations. 


Since electrospray thrusters plume primarily consist of charged droplets, classical methods for plasma simulation in electromagnetic fields were employed for plume modeling, such as Particle-in-Cell (PIC) \cite{emotoNumericalInvestigationSteady2018}. Subsequent studies have explored alternative simulation techniques, including molecular dynamics (MD) simulations\cite{zhang2021molecular, mehta2018sensitivity} and kinetic methods\cite{petroMultiscaleModelingElectrospray2022}. 
However, the fundamental limitation arises because conventional PIC methods are designed for quasi-neutral plasmas, where the cell size must resolve the Debye length to properly capture charge shielding effects. In contrast, electrospray plumes consist of unipolar charged droplets, which lack Debye screening and exhibit strong short-range Coulomb interactions. Moreover, PIC inherently suppresses interparticle Coulomb forces within the same computational cell, making it particularly ill-suited for accurately simulating such highly correlated, non-neutral droplet dynamics.
This fundamental disparity renders conventional plasma simulation approaches potentially inadequate for accurate analysis. Consequently, researchers increasingly adopted Particle-Particle (PP) (also known as n-body or Lagrangian approaches) that explicitly compute Coulomb interactions among individual droplets. Zhao\cite{ZhaoYinjian2019particle-particle, ZhaoYinjian2019particleSimulation} implemented PP simulations to characterize plume development in colloidal thruster acceleration zones. PESPL team developed the DELI model\cite{breddanElectrosprayPlumeDivergence2024, breddanElectrosprayPlumeEvolution2023, breddanElectrosprayPlumeModeling2022, breddanMachineLearningElectrospray2024, wirzElectrosprayThrusterPerformance2019} to systematically investigate background drag effects on electrospray plume evolution. Asher\cite{asherMultiscaleModelingIonic2022} combined a molecular model to investigate the physics of electrospray propulsion. To reduce computational cost, subsequent algorithms introduced cut off far-field Coulomb interactions, though such simplifications risk inducing distortions in electric field distributions and deviations\cite{hampl2022comparison} in particle trajectories. 
While prior studies have developed accurate models for specific aspects, a systematic analysis of the relationship between plume divergence angles and thruster design parameters remains lacking. This study aims to address this gap by investigating how these factors collectively influence thruster lifespan degradation mechanisms.

In this study, a numerical model based on the Particle-Particle (PP) method is developed, achieving high-precision resolution of plume electric fields through particle-wise Coulomb interaction calculations. The research workflow is structured into four phases: First, a simulation model for electrospray plumes is constructed by coupling numerical integration, Coulomb field resolution, and background electric field computation within the PP framework. Subsequently, a sensitivity analysis of plume spatial distribution to key parameters is systematically conducted via controlled variable methods. Following this, the differential effects of Coulomb fields (enhanced in xy- and z-directions) and radial background electric fields on plume expansion are decoupled and investigated. Finally, quantitative relationships between initial parameters and their impacts are derived through fitting, enabling the assessment of critical parameter contributions to plume expansion. This methodology transcends the domain simplification limitations of traditional simulations, providing theoretical support for thruster structural optimization and lifespan prediction.

\section{Simulation method} \label{sec:method}

For charged droplets in an electrospray plume, their motion can be described as:

\begin{equation}
    m \frac{{\rm d}^2 {\textbf{\textit{r}}}}{{\rm d}t^2}
    =m\frac{{\rm d}{\textbf{\textit v}}}{{\rm d} t}
    ={\textbf{\textit{F}}}
    ={\textbf{\textit{F}}}_{\rm pp} + {\textbf{\textit{F}}}_{\rm acc}
    \label{eq:force analysis}
\end{equation}
where $m$ is droplet mass, $t$ is time, $\textbf{r}$ is droplet position vector, $\textbf{v}$ is droplet velocity vector, $\textbf{\textit{F}}$ is the net force acting on the droplet, $\textbf{\textit{F}}_{\rm pp}$ is the particle-particle Coulomb force, $\textbf{\textit{F}}_{\rm acc}$ is the force of background accelerating electric field.
 
 After establishing the electrospray plume model, numerical integration methods can be used in conjunction with boundary conditions to  solve for the background accelerating electric field based on the obtained potential. By substituting the background electric field $\textbf{\textit{E}}_{\rm acc}$ and the inter-droplet interaction force $\textbf{\textit{F}}_{\rm pp}$ calculated according to Coulomb's law into the equation, a physical model for charged droplets in electrospray is established.

\subsection{Numerical integration Methods}
\label{subsec:Numerical integration methods}
Boris pusher is a numerical integration algorithm used for simulating the motion of charged particles in electric and magnetic fields. This algorithm was initially proposed by David J. Boris in 1970\cite{decykAnalyticBorisPusher2023}. It is an efficient algorithm widely applied in plasma simulations. For charged particles, it typically employs the leapfrog scheme\cite{birdsallPlasmaPhysicsComputer1985}, which is generally expressed as:

\begin{equation}
    \left\{ \begin{array}{l}
	\frac{\textbf{x}_{k+1}-\textbf{x}_k}{\Delta t}=\textbf{v}_{k+1/2}\\
	\frac{\textbf{v}_{k+1/2}-\textbf{v}_{k-1/2}}{\Delta t}=\frac{q}{m}\left( \textbf{E}_k+\frac{\textbf{v}_{k+1/2}+\textbf{v}_{k-1/2}}{2}\times \textbf{B}_k \right)\\
\end{array} \right. 
\label{eq:leapfrog}
\end{equation}
In the equation: $\textbf x $ represents the particle position, $\textbf E $ is the background electric field, $ \textbf B $ is the background magnetic field, the subscript $ k $ refers to the quantity at the previous time step, and $ k+1 $ refers to the updated quantity at the next time step (i.e., $ t_k + \Delta t $), and the velocity is calculated between the usual time steps $ t_k $. The Eq.\ref{eq:leapfrog} can be transformed into the form of the Boris pusher as follows:
\begin{equation}
    \left\{ \begin{array}{l}
	\textbf{x}_{k+1}=\textbf{x}_k+\Delta t \textbf{v}_{k+1/2}\\
	\textbf{v}_{k+1/2}=\textbf{v}^+ +q'\textbf{E}_k\\
\end{array} \right. 
\label{eq:boris pusher}
\end{equation}
where
\begin{equation}
    \left\{ \begin{array}{l}
	\textbf{v}^+ =\textbf{v}^- +\left( \textbf{v}^- +\left( \textbf{v}^- \times \textbf{t} \right) \right) \times \textbf{s}\\
	\textbf{v}^- =\textbf{v}_{k-1/2}+q'\textbf{E}_k\\
	\textbf{t}=q'\textbf{B}_k\\
	\textbf{s}=\frac{2{\textbf t}}{ 1+t^2}\\
	q'=\frac{\Delta t q}{2m}\\
\end{array} \right. 
\end{equation}

The steps of the Boris pusher algorithm include: (1) half-time-step acceleration due to the electric field, (2) rotation due to the magnetic field, and (3) half-time-step acceleration due to the electric field. 

The Boris pusher method updates the velocity of each particle by one time step using the electric and magnetic fields obtained for that particle. After acquiring the updated velocity, the position of each particle is then updated to obtain the new position of the particle after one time step.

The leap frog method with Boris algorithm is predominantly employed in plasma simulations due to its proficiency in handling electromagnetic field challenges. Since electrospray thrusters typically operate without magnetic fields, the velocity Verlet method has also been investigated as an alternative pusher algorithm, given that both approaches maintain second-order accuracy. Consequently, under conditions that include electric and magnetic fields, these two numerical integration methods may exhibit distinct computational performances. In the following sections, a comparative analysis of the computational effectiveness of these methods will be presented in \ref{subsec:integration comparison}.

Velocity Verlet method is typically utilized in the fields of molecular dynamics and Monte Carlo simulations to compute the evolution of particles over time, thereby obtaining information such as particle trajectories and velocities. It is important to note that the performance and stability of the velocity Verlet method depend on the choice of time step; a smaller time step generally enhances the accuracy of the simulation but also increases computational cost. Therefore, when employing this method, it is necessary to select an appropriate time step.

Unlike the Boris pusher, in the velocity Verlet algorithm, the velocity and position of particles are calculated at the same value of time variable, as shown in Eq.\ref{eq:Verlocity Verlet Equation}, which explicitly incorporates velocity, solving the issue of the first time step in the basic Verlet algorithm.

\begin{equation}
    \left\{ \begin{array}{l}
        \textbf{x}\left( x+\Delta t \right) =\textbf{x}\left( t \right) +\textbf{v}\left( t \right) \Delta t+\frac{1}{2}\textbf{a}\left( t \right) \Delta t^2\\
	\textbf{v}\left( t+\Delta t \right) =\textbf{v}\left( t \right) +\frac{\textbf{a}\left( t \right) +\textbf{a}\mathrm{(t+\Delta t)}}{2}\Delta t\\
    \end{array} \right. 
\label{eq:Verlocity Verlet Equation}
\end{equation}

The KDK form of velocity Verlet is demonstrated in Eq.\ref{eq:Verlocity Verlet after}.

\begin{equation}
    \left\{ \begin{array}{l}
    \textbf{x}({{t+\Delta t}})=\textbf{x}({t})+\textbf{v}({t})\Delta t+\frac{1}{2}\textbf{a}({t})\Delta t^2\\
	\textbf{F}=m\textbf{a}({t+\Delta t})\\
	\textbf{v}({t+\Delta t})=\textbf{v}({t})+\textbf{v}({t})\Delta t+\frac{1}{2}\left( \textbf{a}({t})+\textbf{a}({t+\Delta t}) \right) \Delta t^2\\

    \end{array} \right. 
\label{eq:Verlocity Verlet after}
\end{equation}

Compared to other numerical integration methods, velocity Verlet method explicitly updates the velocity at each step, allowing for a more accurate calculation of the system's energy. Additionally, the algorithm is straightforward and easy to implement in computer programs, making it suitable for long-term dynamical simulations with minimal energy drift.

\subsection{Particle-particle Coulomb field}
\label{subsec:pp field}

The Coulomb electric field exerted on the $i$-th droplet by all other droplets (the electric field force generated by charged droplets on charged droplets) is calculated using Coulomb's law:

\begin{equation}
    \textbf{\textit{E}}_i = \frac{\textbf{\textit{F}}_i}{q_i}
    =\frac{1}{4\pi \varepsilon_0}\sum^N_{j=1,j\ne i}  \frac{q_j}{|\textbf{\textit{r}}_{ij}^3|}{\textbf{\textit{r}}_{ij}}
\end{equation}
where $\textbf{\textit{r}}_{ij}$ is the distance vector from the $i$-th droplet to the $j$-th droplet, $N$ is the total number of simulated droplets. By calculating the distance between the two droplets, it computes the Coulomb electric field for every simulated droplets, and cutoffs are not needed for all like charge system.


\subsection{Accelerating electric field}
The working principle of an electrospray thruster involves applying a high voltage between the emitter and extractor electrodes to create an electric field\cite{ZhaoYinjian2019particleSimulation}. This electric field extracts and accelerates charged particles from the ionic liquid, thereby generating thrust. Consequently, in the simulation process, it is essential to accurately model the background acceleration electric field formed between the emitter and extractor electrodes. Since the focus of this study is on analyzing the mutual Coulomb interactions between charged droplets, the simulation of the acceleration electric field is achieved by solving Poisson's equation to obtain its analytical solution.The Poisson equation is frequently employed to describe the distribution of electric potential or other potential fields in physical systems.In a three-dimensional cylindrical coordinate system, it takes the form as shown in Eq.\ref{eq:Poission}, where $ r $, $ \theta $, and $ z $ represent the radial, circumferential, and axial coordinates, respectively.
\begin{equation}
    \nabla ^2\phi =\frac{\partial ^2\phi}{\partial r^2}+ \frac{1}{r}\frac{\partial \phi}{\partial r} +\frac{1}{r^2}\frac{\partial ^2\phi}{\partial \theta ^2}+\frac{\partial ^2\phi}{\partial z^2}=-\frac{\rho}{\varepsilon _0}
    \label{eq:Poission}
\end{equation}
Subsequently, the prolate spheroidal coordinates $(\xi, \eta, \phi)$ are used, and the coordinate transformation is carried out using Eq.\ref{eq:prolate spheroidal transformation}.
\begin{equation}
    \left\{ \begin{array}{l}
	\zeta =\frac{r_1+r_2}{a}\\
	\eta =\frac{r_1-r_2}{a}\\
	r_1=\sqrt{x^2+y^2+\left( z+\frac{a}{2} \right) ^2}\\
	r_2=\sqrt{x^2+y^2+\left( z-\frac{a}{2} \right) ^2}\\
\end{array} \right. 
    \label{eq:prolate spheroidal transformation}
\end{equation}

\begin{figure}
    \centering
    \includegraphics[width=0.55\linewidth]{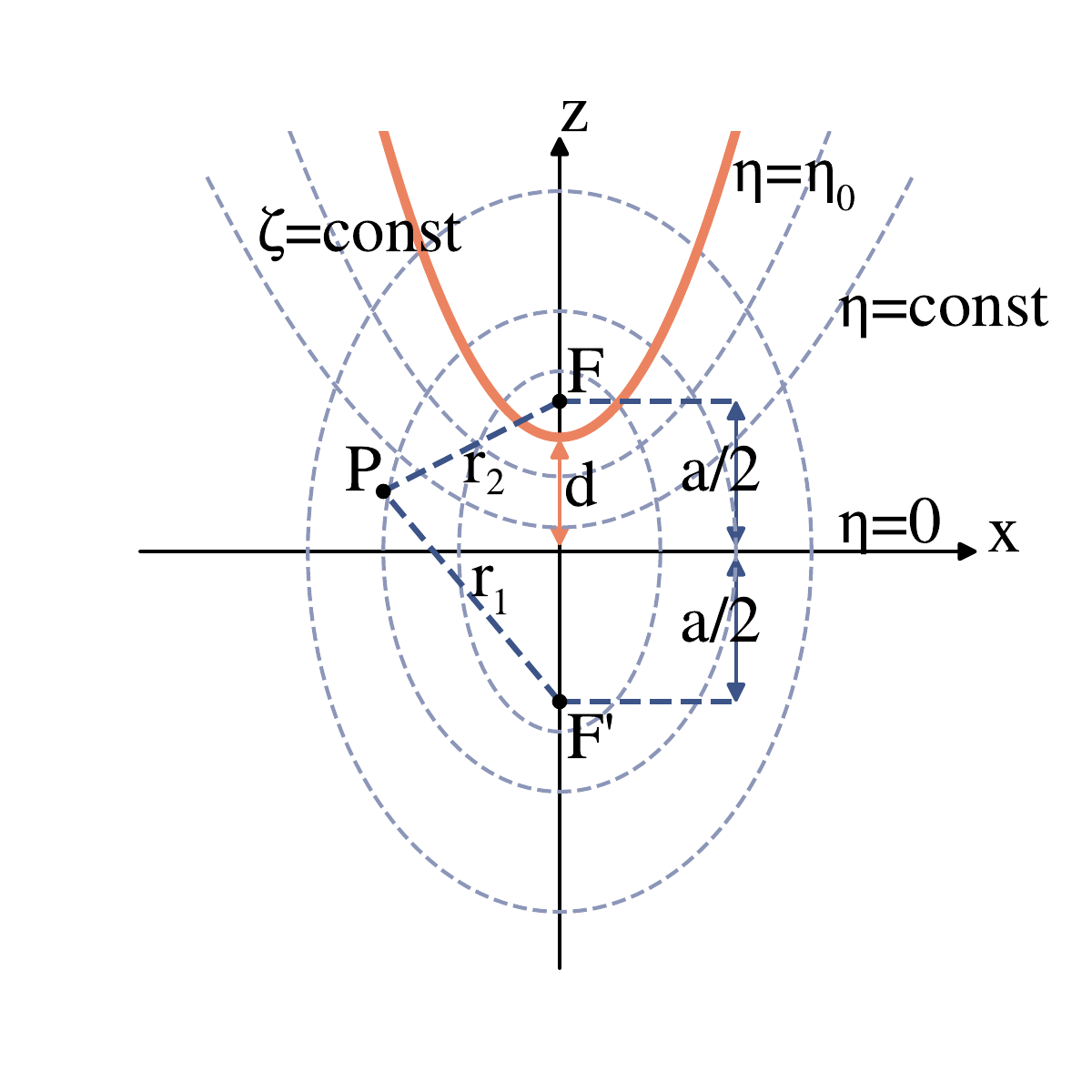}
    \caption{The prolate spheroidal coordinate system.}
    \label{fig:prolate spheroidal coordinate system}
\end{figure}

As shown in Fig.\ref{fig:prolate spheroidal coordinate system}, where $ a $ is the semi-focal distance of the hyperboloid, and $ \phi $ is the azimuthal angle with respect to the line FF' (not shown in the figure) and the y-axis pointing towards the paper. Lines of constant $ \xi $ are confocal ellipsoids with the same foci (F and F'), while lines of constant $ \eta $ are confocal hyperboloids. In the simulation model, the surface $ \eta = 0 $ represents the acceleration electrode plane, the surface $ \eta = \eta_0 $ represents the liquid surface, i.e., the Taylor cone, and $ d $ is the vertical distance between the tip of the liquid surface and the acceleration plate.
In order to simplify the calculation, let's assume that the potential of the plate (acceleration electrode) is 0, and the potential at the Taylor cone liquid surface is $\phi_0$. Then, the Laplace equation for the background acceleration electric field distribution at the Taylor cone in the simulation area depends only on the $\eta$ in the space. Therefore, the Laplace equation for the background acceleration electric field at the Taylor cone is as Eq.\ref{eq:accelerateing field Lapalce equation}.
\begin{equation}
    \frac{\partial}{\partial \eta}\left[ \left( 1-\eta ^2 \right) \frac{\partial \phi _{\mathrm{acc}}}{\partial \eta} \right] =0
    \label{eq:accelerateing field Lapalce equation}
\end{equation}
Combining the two boundary conditions that $\eta = \eta_0$ at the liquid surface and $\eta=0$  at the plate, we can determine the potential distribution of the acceleration electric field. Then, we can use the Poisson equation to obtain the spatial distribution of the electric field, as shown in Eq.\ref{eq:spatial distribution}.
\begin{equation}
    \textbf{E}_{\mathrm{acc}}=-\nabla\phi_{\mathrm{acc}}=\left( -\frac{\partial \phi _{\mathrm{acc}}}{\partial \eta}\frac{\partial \eta}{\partial x},-\frac{\partial \phi _{\mathrm{acc}}}{\partial \eta}\frac{\partial \eta}{\partial y},-\frac{\partial \phi _{\mathrm{acc}}}{\partial \eta}\frac{\partial \eta}{\partial z} \right) 
    \label{eq:spatial distribution}
\end{equation}

Therefore, the electric field distribution can be calculated as long as we know the values of a and $\eta_0$. Considering that at the tip of the Taylor cone, we have $x=0$, $y=0$, $z=d$, and furthermore $\eta=\frac{2d}{a}=\eta_0$, we can derive Eq.\ref{eq:Taylor cone tip}.
\begin{equation}
    \left\{ \begin{array}{l}
	a=2d\left( 1+\frac{R_c}{d} \right) ^{\frac{1}{2}}\\
	\eta _0=\left( 1+\frac{R_c}{d} \right) ^{-\frac{1}{2}}\\
\end{array} \right. 
    \label{eq:Taylor cone tip}
\end{equation}

Determining $R_c$ and $d$ is sufficient to establish the values of $a$ and $\eta_0$. At the electrospray nozzle, the geometric relationship between the nozzle radius and the Taylor cone results in the relationship between the nozzle radius $R_n$ and the Taylor cone radius $R_c$, as given by Eq.\ref{eq:Rc and Rn}.
\begin{equation}
    R_{\mathrm{c}}=\frac{\mathrm{R}_{\mathrm{n}}}{\cos \theta _{\mathrm{T}}}
    \label{eq:Rc and Rn}
\end{equation}

Where $\theta_T = 49.29^\circ$, it is thus possible to solve for the background acceleration electric field of the Taylor cone using Poisson's equation.

\section{Validation} 
\label{sec:validation}

\subsection{Numerical integration method comparison}
\label{subsec:integration comparison}
In a uniform electric field with magnetic field, a particle with a certain charge and mass has its theoretical trajectory described as follows\cite{chenIntroductionPlasmaPhysics2016}
\begin{equation}
    \left\{ \begin{array}{l}
	x-x_0=r_L\sin \omega _ct\\
	y-y_0=\pm r_L\cos \omega _ct\\
	\omega _c\equiv\frac{|q|B}{m} \\
	r_L=\frac{v_z}{\omega _c}\\
\end{array} \right. 
\label{eq:theoretical trajectory}
\end{equation}
In the equation, $(x, y)$ represents the position coordinates of the charged particle at a certain moment, $(x_0, y_0)$ represents the initial position coordinates of the charged particle (in meters), $\omega_c$ is the cyclotron frequency of the charged particle; $r_L$ is the Larmor radius of the charged particle. This serves as the theoretical value for the motion of charged particles in an electrtic field with magnetic field, and the simulated values obtained by the Boris pusher and velocity Verlet numerical integration methodd are compared with it to conduct error tests of these two numerical integration algorithms. The tests employ the model particle parameters shown in Tab.\ref{tab:parameters for Boris}.

\begin{table}[htbp]
	\centering
	\begin{tabular}{cccc}
		\hline
		Parameter & Symbol & Value & Unit       \\ \hline
		Droplet charge&$q$        & 1.0 &   C     \\
		Droplet mass&$m$        & 1.0  &    kg   \\
		Magnetic field&$\textbf B$       & 0, 0, 1.0 & T \\
		Electric field&$\textbf E$        & 0, 0, 1.0 & V/m\\
		Droplet initial velocity &$\textbf v_0$      & 0, 0, 0  & m/s \\
		Time step&$\Delta t$  & 0.05 &  s     \\
		Total simulation time&$t$        & 20  &    s    \\ \hline
	\end{tabular}
	\caption{Parameters for Boris pusher simulation error test.}
	\label{tab:parameters for Boris}
\end{table}

\begin{figure}[htbp]
	\centering
	\includegraphics[width=0.9\linewidth]{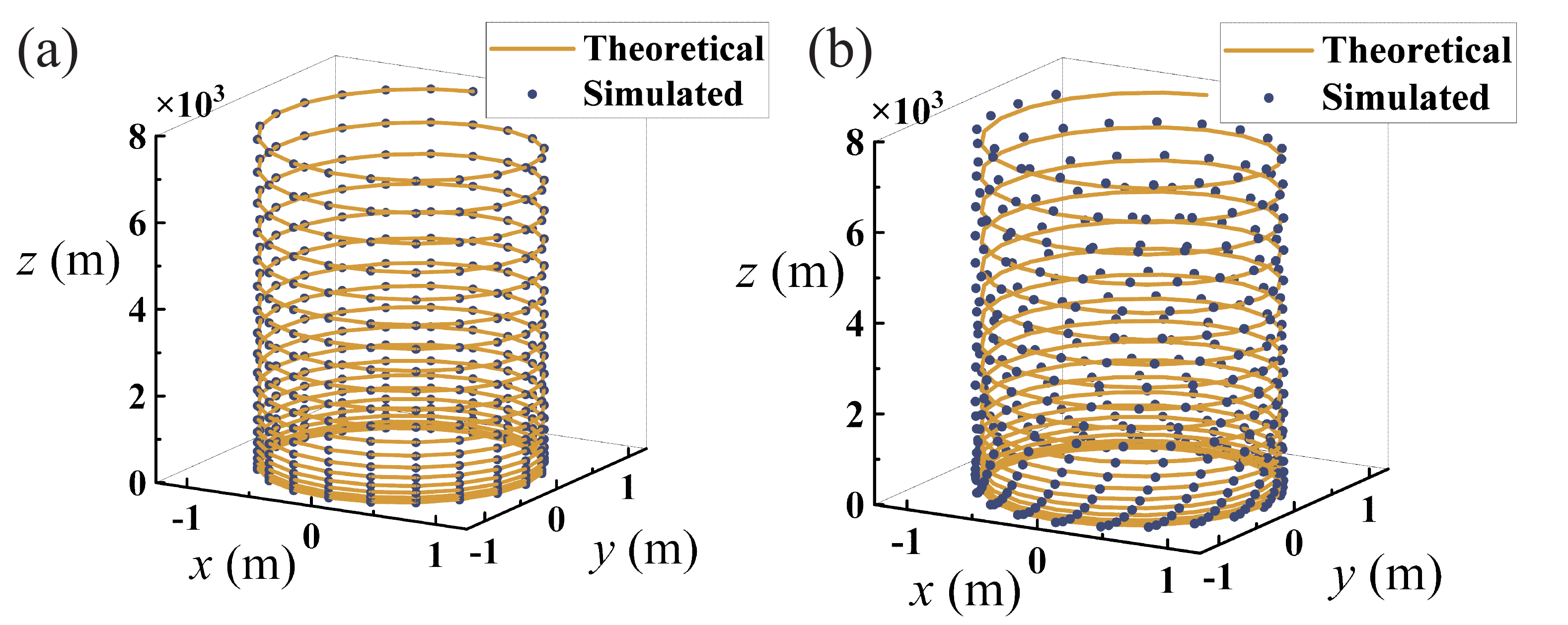}
	\caption{Comparison Between Theoretical and Boris pusher Simulated Trajectories of Charged Particle Motion.}
	\label{fig:Boris test}
\end{figure}

Using model charge droplet parameters for case error testing, the test trajectories are shown in Fig.\ref{fig:Boris test}. In this figure, the ``$-$'' represents the theoretical trajectory of the charged droplet, while the ``\ $\cdot$\ '' indicates the simulated trajectory of the charged droplet using Boris pusher method and velocity Verlet method. The Boris pusher simulation result is relatively accurate, while the velocity Verlet simulation result appears to be less precise.

\begin{figure}[htbp]
\centering
\includegraphics[width=1.0\linewidth]{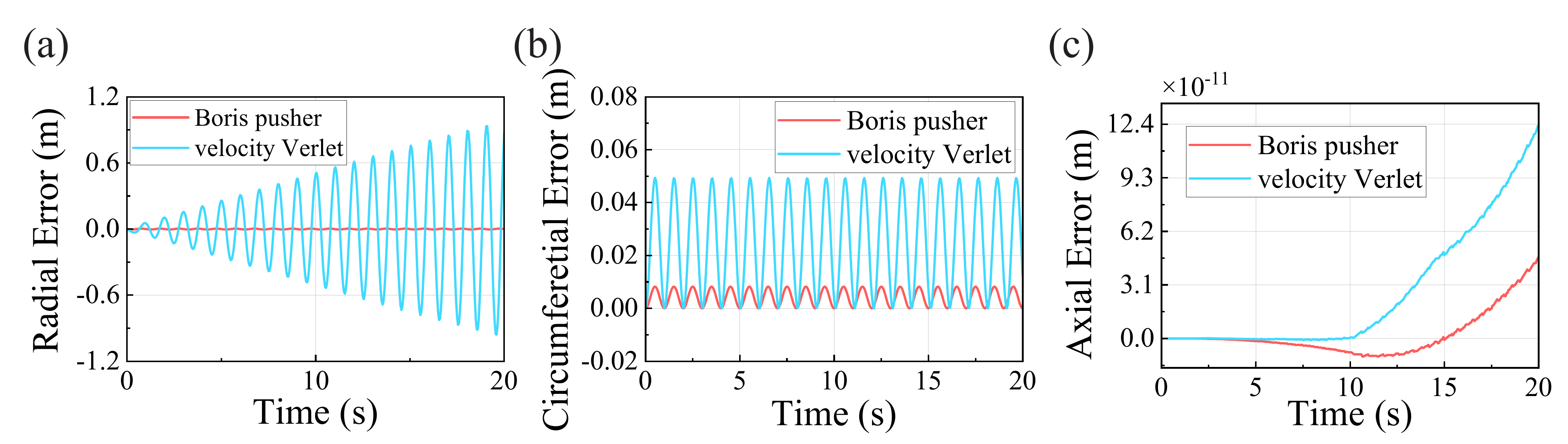}
\caption{
Error analysis in charged particle trajectory simulations using the boris pusher algorithm. Figures (a), (b), and (c) respectively represent the radial error, circumferential error, and axial error during the simulation process.
}
\label{fig:boris test error}
\end{figure}

Fig.\ref{fig:boris test error} shows the simulation error of particle motion using the Boris pusher method and velocity Verlet method. Fig.\ref{fig:boris test error}(a) represents the radial error. Boris pusher consistently maintains its radial error within a range of $\pm 0.0045$ m over a relatively short period of time. In contrast, velocity Verlet initially exhibits a smaller radial error, but as time progresses, the error accumulates, reaching a range of $\pm 1.0$ m by the end of 20 seconds. In terms of radial error, the error of Boris pusher is significantly smaller than that of velocity Verlet. As depicted in Fig.\ref{fig:boris test error}(b), regarding azimuthal error, both Boris pusher and velocity Verlet maintain their azimuthal errors within certain limits over a short period of time. However, Boris pusher consistently keeps its error within $0.0009$ m, while the error of velocity Verlet reaches $0.05$ m. In terms of azimuthal error, the error of Boris pusher is significantly smaller than that of velocity Verlet. Fig.\ref{fig:boris test error}(c) illustrates the axial error, both Boris pusher and velocity Verlet maintain their axial errors within a relatively small range over a short period of time, with the magnitude being on the order of $10^{-11}$-$10^{-10}$ m. The error is small and can be neglected. In terms of axial error, both Boris pusher and velocity Verlet exhibit small errors, which are within an acceptable range.

Overall, when simulating with model particle parameters, the Boris pusher demonstrates smaller errors in radial, azimuthal, and axial directions compared to the velocity Verlet method. It exhibits superior computational performance when electric field and magnetic field are involved in the simulation. Therefore, the Boris pusher method will be employed as the numerical integration method in subsequent simulations.

\subsection{Coulomb binary collision test}
\label{subsec: coulomb test}

\begin{figure}[htbp]
    \centering
    \includegraphics[width=0.5\textwidth]{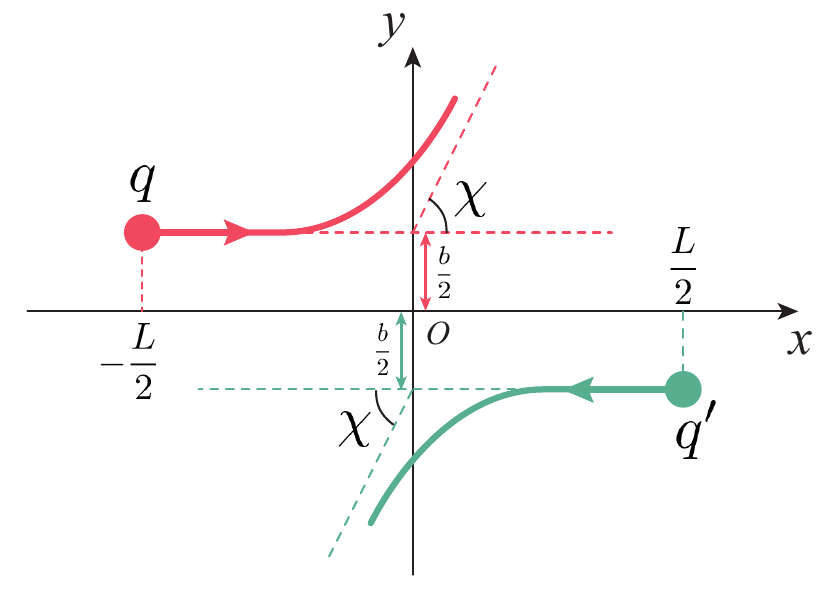}
    \caption{Binary collision setup.}
    \label{fig:binary_collision_setup}
\end{figure}

By simulating binary collisions to test the accuracy of the Coulomb field algorithm. As shown in Fig.\ref{fig:binary_collision_setup}, two droplets on the same plane move towards each other with a certain initial velocity. 

The initial coordinates of the droplets are set as $(-\frac{L}{2},\frac{b}{2})$ and $(\frac{L}{2},-\frac{b}{2})$, and the angle with the x-axis after collision is denoted as $\chi$. According to the binary collision theory\cite{robinsonBinaryCollisionApproximation1994}, the theoretical scattering angle $\chi$ is given by Eq.\ref{eq:scattering angle}.

\begin{equation}
    \left\{ \begin{array}{l}
	\tan \frac{\chi}{2}=\frac{b_0}{b}\\
	b_0=\frac{1}{4\pi \varepsilon _0}\frac{qq'}{\mu g^2}\\
\end{array} \right. 
\label{eq:scattering angle}
\end{equation}
In the formula, $g$ represents the sum of the velocities of the two droplets moving towards each other, $b$ represents the longitudinal distance between the two droplets, $q$ and $q'$ represent the charges of the two droplets (which are equal in this test), $m$ and $m'$  represent the masses of the two droplets (which are equal in this test), $\mu$ represents the reduced mass of the two droplets during the collision, which is calculated as 
\begin{equation}
    \mu =\frac{mm'}{\left( m+m' \right)}
\end{equation}

\begin{figure}[htbp]
\centering
\includegraphics[width=1.0\textwidth]
{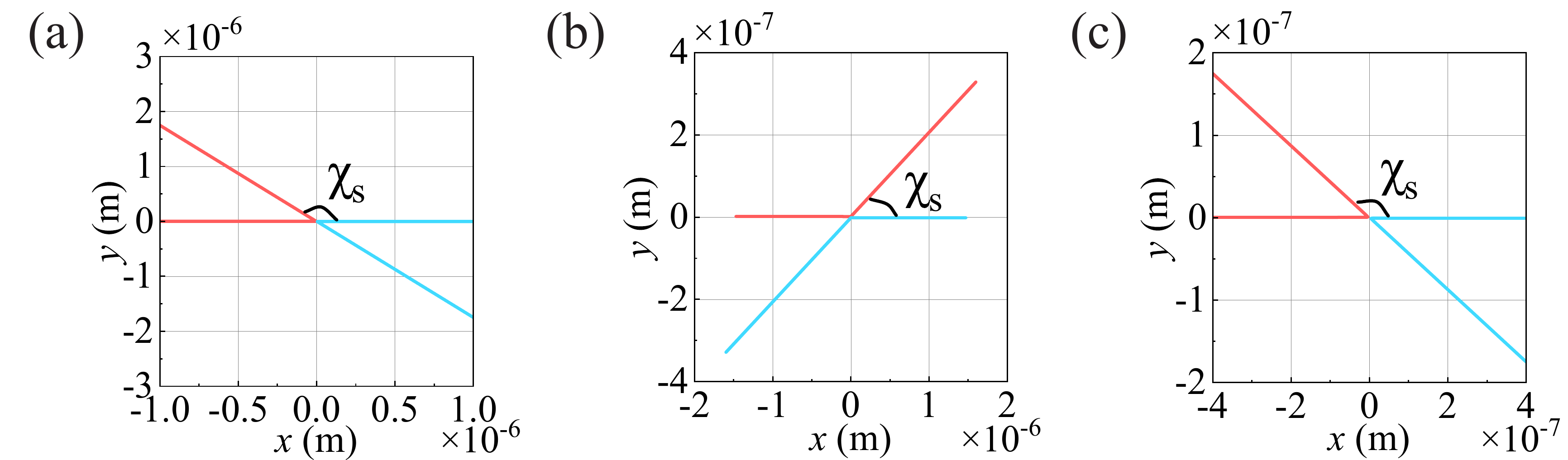}
\caption{
Trajectory of droplets of binary collision similation error test. (a) Test 1. (b) Test 2. (c) Test 3.
}
\label{fig:Coulomb test error}
\end{figure}

The numerical integration algorithm used in the test is the Boris pusher. Three test cases were conducted, and the parameters and results are shown in Tab.\ref{tab:Parameters for binary collision simulation test}. The trajectories of the droplets during the simulation are shown in Fig.\ref{fig:Coulomb test error}. From the errors, it can be observed that the Coulomb field algorithm has a relatively small error, and as the time step decreases, the error also tends to decrease. Therefore, within a certain time step, the error of the Coulomb field algorithm is acceptable.

\begin{table}[htbp]
	\centering
	\begin{tabular}{cccccc}
		\hline
		   Parameter & Symbol     & Test 1                     & Test 2                     & Test 3      & Unit               \\ \hline
	Droplet charge	& $q$            & $3.77\times 10^{-18}$ & $2.36\times 10^{-19}$ & $1.60\times 10^{-19}$ & C\\
	Droplet mass	& $m$            & $8.15\times 10^{-22}$ & $5.09\times 10^{-23}$ & $1.85\times 10^{-25}$ & kg \\
		Velocity sum & $g$            & 257                   & 257                   & 751                  & m/s \\
		Longitudinal distance & $b$            & $2.94\times 10^{-9}$  & $2.94\times 10^{-9}$  & $0.92\times 10^{-9}$ & m  \\
		Time step & $\Delta t$     & $5.15\times 10^{-11}$ & $3.22\times 10^{-13}$ & $1.28\times 10^{-13}$ & s \\
		Theoretical angle & $\chi_t$       & 2.0323                & 0.2018                & 2.7302     & rad           \\
		Simulated  angle &$\chi_s$       & 2.0898                & 0.2018                & 2.7290       &rad         \\
		Relative Error & $E_r$ & 2.8261\%               & -0.0004\%              & -0.0431\%  & -            \\ \hline
	\end{tabular}
	\caption{Parameters for binary collision simulation test}
	\label{tab:Parameters for binary collision simulation test}
\end{table}

\subsection{Accelerating electric field test}
Along the $y=y_0$ direction, the $x-z$ plane is divided into a grid of $1000\times 1000$ points between the electrospray exit and the collector plate. Based on Eq.\ref{eq:spatial distribution} and Eq.\ref{eq:Taylor cone tip}, an outer and inner loop is used to 
perform calculations in the $x$ and $z$ directions, respectively, to obtain the Taylor cone background acceleration electric field at each of the $1000\times 1000$ grid points. The data is then output and saved.

\begin{figure}[htbp]
\centering
\includegraphics[width=1.0\linewidth]
{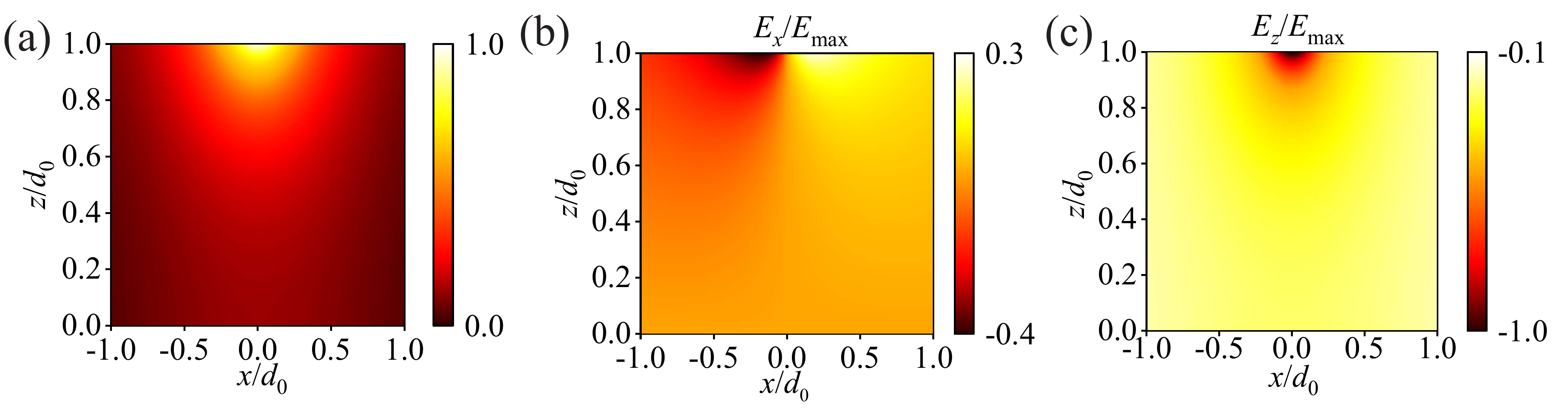}

\caption{The figure presents the simulation results for the electric potential distribution and electric field distribution in a selected region. Figure (a) represents the electric potential distribution, while figures (b) and (c) depict the radial and axial electric field distributions, respectively.
}
\label{fig:accelerating field}
\end{figure}

After setting the parameter $R_n/d=0.2$ and $y=0$ for simulation, the potential and electric field simulation results are shown in Fig.\ref{fig:accelerating field}. $\phi_0$ is the voltage drop from the tip of the Taylor cone to the acceleration plate. Correspondingly, the obtained data is normalized , setting  such that the data is multiples of $E_{\mathrm{max}}$. 

\section{Results and discussion}
\label{sec: results and discussion}

\subsection{Simulation domain and parameters}
\begin{figure}[htbp]
    \centering
    \includegraphics[width=0.5\linewidth]{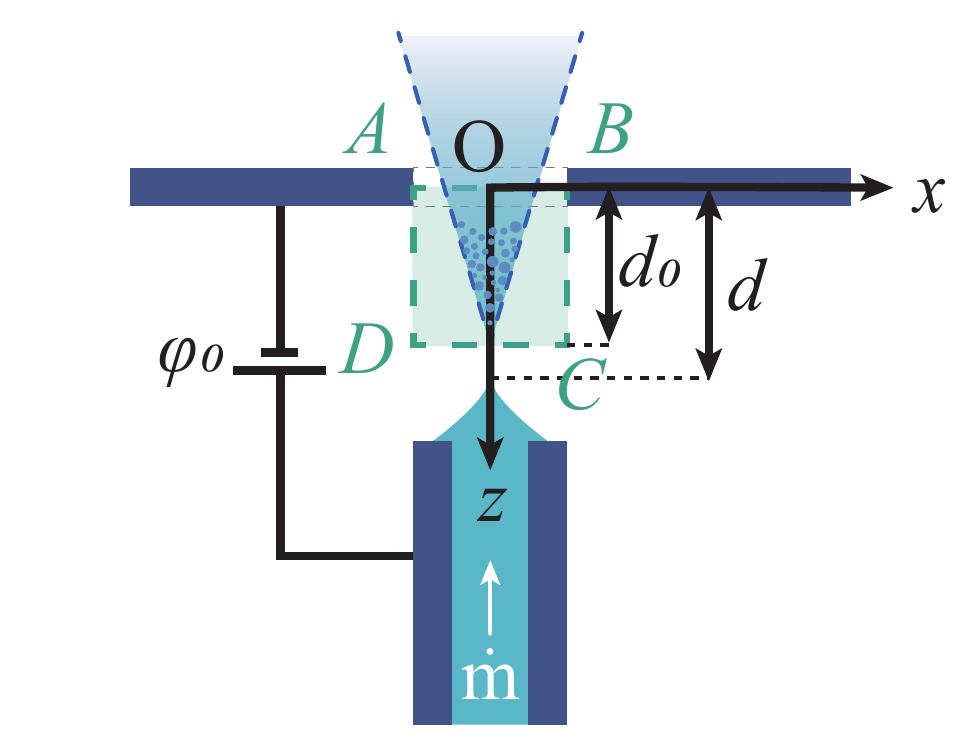}
    \caption{Simulation domain is the green shaded region ABCD in the figure.}
    \label{fig:simulation domain}
\end{figure}

The model of an electrospray thruster can be simplified as depicted in Fig.\ref{fig:simulation domain}, where the spray needle and the accelerating electrode (represented in gray) are connected to a power supply with a potential difference of $\phi_0$. The spray needle (depicted in blue) contains an electrospray liquid. Due to the potential difference, i.e., the electric field acting on the liquid surface, a Taylor cone is formed at the tip of the needle, and a jet is emitted from the tip of the cone\cite{ZhaoYinjian2019particle-particle}. At a certain moment, the jet decomposes into droplets, and the larger droplets further decompose into smaller droplets, which are all accelerated towards the accelerating electrode.

As shown in the green shaded area ABCD of Fig.\ref{fig:simulation domain}, the simulation region is selected at the position where the jet breaks and accelerates towards the acceleration plate. The origin of the simulation domain is chosen as the point O marked in the figure; the z-axis is perpendicular to the acceleration plate, pointing along the symmetry axis of the needle; the x-axis is along the direction of the flat plate; the y-axis is perpendicular to the paper surface pointing inward (not shown in the figure). Thus, the simulation region is a three-dimensional cube that encompasses all droplets. The distance from the tip of the Taylor cone to the acceleration plate is denoted by $d$, and the distance from the jet break position to the acceleration plate is denoted by $d_0$. In the numerical simulation of this paper, droplets are allowed to move radially to any position away from the z-axis, meaning there are no boundaries in the simulation domain in the x and y directions.

\begin{figure}[htbp]
    \centering
    \includegraphics[width=0.4\linewidth]{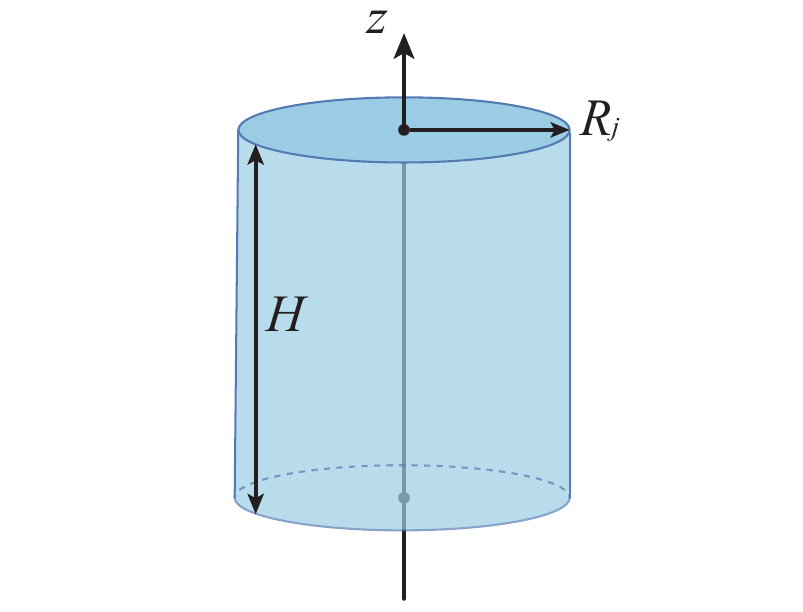}
    \caption{The injection volume within a single time step.}
    \label{fig:injection volume}
\end{figure}

The simulation begins at the onset of jet breakup, necessitating the placement of an appropriate volume and number of initial droplets at the front end of the jet to simulate the emission of droplets. If the time step in the numerical simulation is denoted by $\Delta t$, then the injection volume within the time step can be correspondingly determined. As shown in Fig.\ref{fig:injection volume}, the corresponding injection volume is a cylinder with a radius $R_j$ and height $H$. Point A is located at $(0,0,d_0)$ within the simulation region, with the cylinder's axis being the z-axis. The position of the droplets emitted by injection within the injection volume is random, meaning the initial coordinates of each droplet are determined using the Box-Muller method\cite{dahmaniEfficientFPGABasedGaussian2023}. For each droplet, its initial velocity is composed of the thermal velocity $v_t$ caused by the droplet's own temperature and the drift velocity $v_d$ brought by the jet. In this simulation, the time step is set to no more than $1\%$ of the emission time and is the closest integer unit time length. For instance, when the emission time interval is $2.678 \times 10^{-10}$ seconds, the time step will be set to $1 \times 10^{-12}$ s to reduce simulation error.

The geometric parameters, material parameters, and control parameters of the electrospray thruster will refer to the data obtained by Miller\cite{millerCapillaryIonicLiquid2021} using mass spectrometry. The working fluid uses EMI-IM with a flow rate of 1.31 nl/s, and the simulation mainly uses the parameters related to the large droplets from the data. The charge of a single droplet is $ q = 1.442 \times 10^{-17} $ C, the mass of a single droplet is $ 5.99 \times 10^7 $ amu, the initial drift velocity $ v_d = 581.9 $ m/s, the distance from the tip of the jet to the acceleration plate $ d_0 = 0.0015 $ m, and the acceleration voltage is 750 V.


\subsection{Electrospray plume shape}
\label{subsec: plume shape}

\begin{figure}[htbp]
    \centering
    \includegraphics[width=0.8\linewidth]{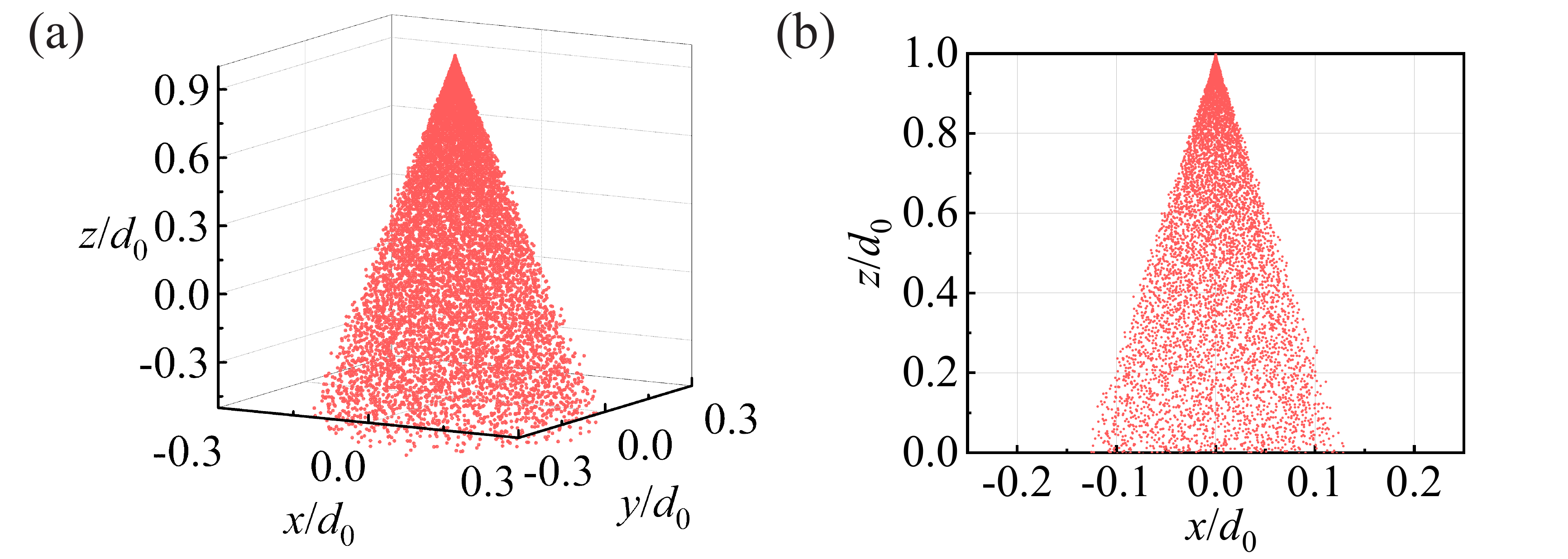}
    \caption{The figure depicts the simulated shape of the electrospray thruster plume, which includes (a) the spatial distribution of the electrospray plume, and (b) the x-z phase space image of the electrospray plume.}
    \label{fig:plume shape}
\end{figure}

Fig.\ref{fig:plume shape}(a) presents a three-dimensional spatial distribution of the electrospray plume, while Fig.\ref{fig:plume shape}(b) illustrates the x-z plane distribution of the electrospray droplets. The droplets are ejected from the spray needle and jet, originating at the position $(0,0,d_0)$, and are accelerated in the -z direction while expanding radially. It can be observed in the figure that the electrospray plume gradually expands after being emitted.

\subsection{Droplets velocity analysis}
\label{subsec: droplets velocity}
\begin{figure}
    \centering
    \includegraphics[width=1.0\linewidth]{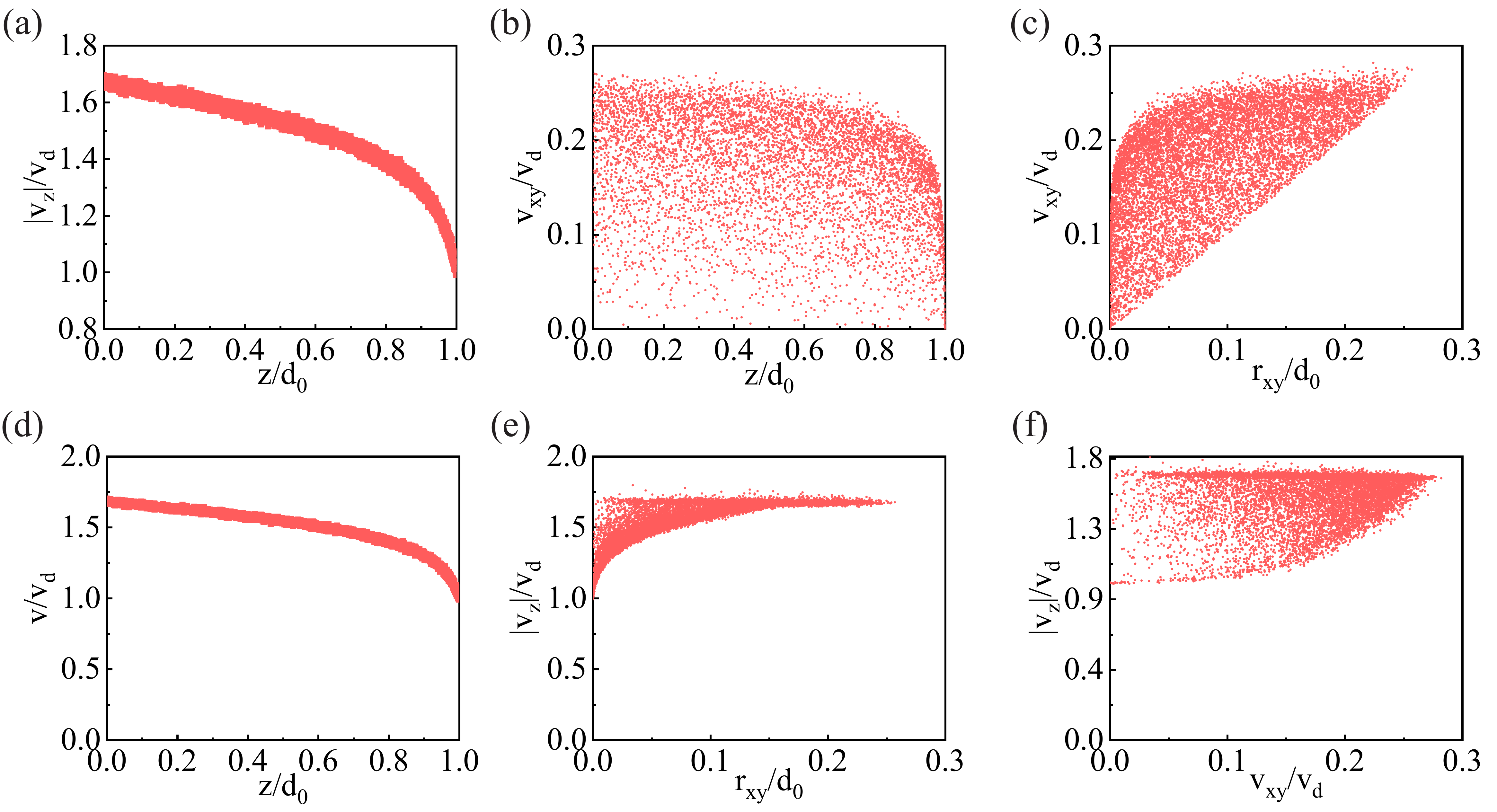}
    \caption{Phase space diagram of droplet velocities in the electrospray plume.}
    \label{fig:velocity phase plot}
\end{figure}
Fig.\ref{fig:velocity phase plot}(a) presents the $z$-$v_z$ phase space diagram of the droplets in the electrospray plume. All droplets are accelerated from $ v_d $ in the negative $z$-direction to approximately $ 1.7v_d $ at the final acceleration plate. Therefore, the simulated thrust of a single spray needle electrospray thruster is approximately $ F_T = 1.7v_d\rho\dot{V} \approx 1.97 \times 10^{-6} $ N, and the specific impulse is approximately $ I_{sp} = 1.7v_d/g_0 \approx 101 $ s.

Fig.\ref{fig:velocity phase plot}(b) presents the $z$-$v_{xy}$ phase space diagram of the droplets in the electrospray plume, where $ v_{xy} = \sqrt{v_x^2 + v_y^2} $. After being ejected from the spray needle, the droplets first experience a significant acceleration due to the electric field force and particle-particle Coulomb force, causing their radial velocity to gradually increase. As the droplets approach the acceleration plate, their radial acceleration decreases, resulting in a diminishing change in radial velocity.

Fig.\ref{fig:velocity phase plot}(c) presents the $r_{xy}$-$v_{xy}$ phase space diagram of the droplets, where $ r_{xy} = \sqrt{x^2 + y^2} $. A distinct diagonal line is visible in the diagram, indicating that only droplets with higher $ v_{xy} $ can reach larger $ r_{xy} $, which corresponds to the outer regions of the plume; whereas droplets with lower $ v_{xy} $ remain near $ r_{xy} = 0 $, which corresponds to the inner regions of the plume. Finally, Fig.\ref{fig:velocity phase plot}(d) shows the z-v diagram of the droplets, Fig.\ref{fig:velocity phase plot}(e) presents the $r_{xy}$-$v_z$ diagram, and Fig.\ref{fig:velocity phase plot}(f) illustrates the $v_{xy}$-$v_z$ diagram.

\subsection{Influence of key factors}
\label{subsec: key factors}

To explore the effects of various parameters, including material and operational parameters, on the distribution of electrospray plumes, the numerical simulation involves altering the value of a single parameter and calculating the half-angle of the electrospray plume. This assessment determines the impact of the modified parameter on the plume's half-angle. 
\begin{equation}
    \theta =\arctan \frac{\sqrt{x^2+y^2}}{z}
    \label{eq:plume angle}
\end{equation}
Eq.\ref{eq:plume angle} is utilized to calculate the half-angle of the electrospray plume, focusing on the droplets within the range of $0 \leq z \leq 0.05d_0$. Subsequently, the maximum plume half-angle (100th percentile value), 90th percentile value, and 80th percentile value obtained from the simulation region are selected as the results. Corresponding diagrams illustrating the relationship between the parameters and the plume half-angle are then created.

\begin{figure}
    \centering
    \includegraphics[width=1.0\linewidth]{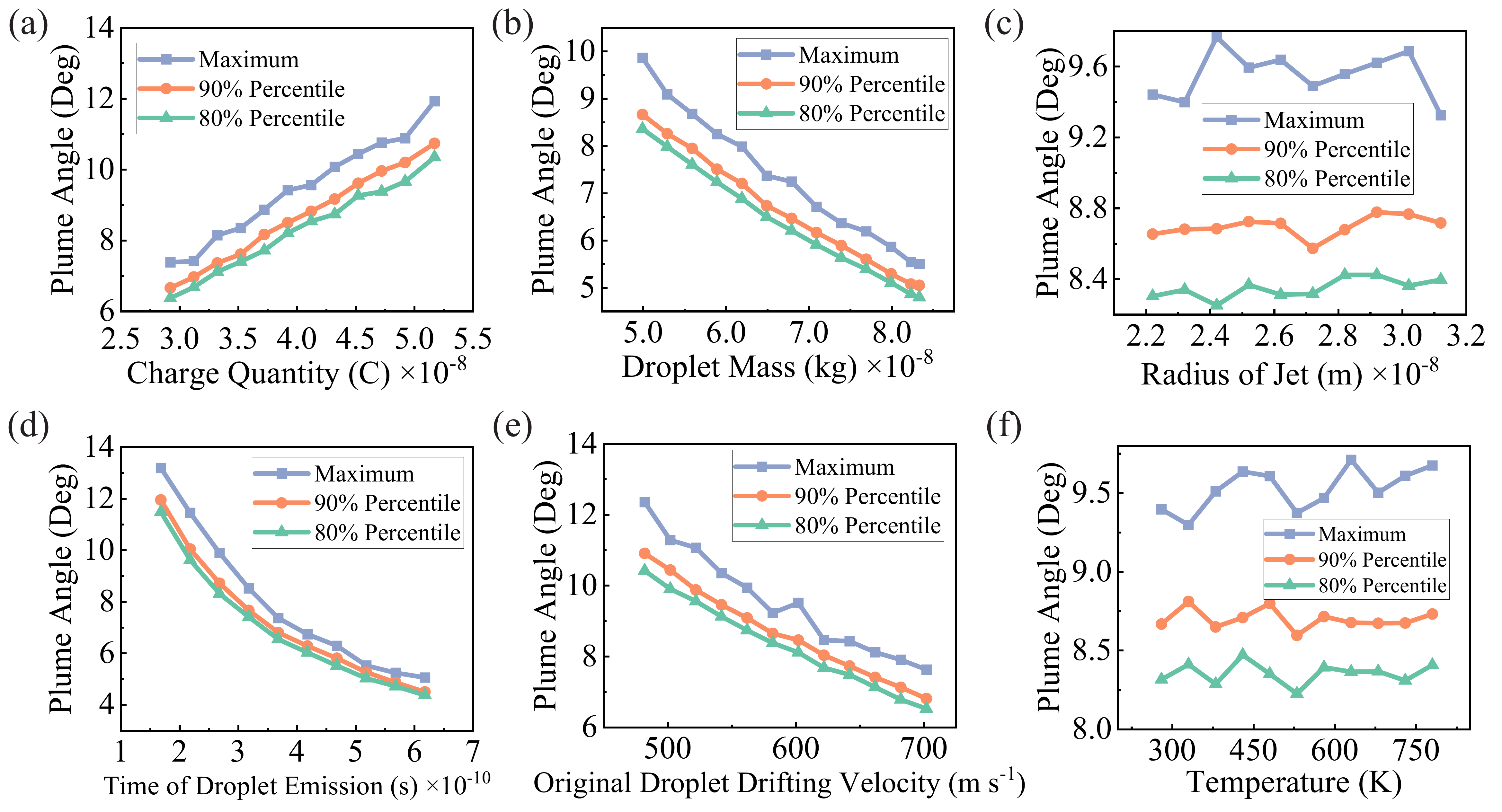}
    \caption{The impact of various parameters on the half-angle of the plume.}
    \label{fig:plume angle - parammeters}
\end{figure}

As shown in Fig.\ref{fig:plume angle - parammeters}, this study investigates a total of six simulation parameters. Among them, Fig.\ref{fig:plume angle - parammeters}(a) explores the parameter of droplet charge quantity. As the charge quantity increases, the Coulomb interaction between droplets is enhanced, leading to a significant increase in the half-angle of the electrospray plume. Fig.\ref{fig:plume angle - parammeters}(b) explores the parameter of droplet mass. As the mass increases, the acceleration produced by a certain electric field force decreases, the expansion effect of the plume is reduced, and the half-angle of the electrospray plume is significantly decreased. Fig.\ref{fig:plume angle - parammeters}(d) explores the parameter of droplet emission time interval. As the emission time interval increases, the number of droplets (density) in the simulation area within the same period of time decreases, the Coulomb interaction between droplets is weakened, and the half-angle of the electrospray plume is significantly reduced. Fig.\ref{fig:plume angle - parammeters}(e) explores the parameter of the initial drift velocity of droplets. As the initial drift velocity increases, the half-angle of the electrospray plume is significantly decreased.

Fig.\ref{fig:plume angle - parammeters}(c) and (f) investigate the jet radius and temperature, respectively, and no clear trends are observed. It is analyzed that the jet radius may only affect the initial position of the droplets. Since the jet radius is of a smaller order of magnitude, minor changes may not affect the plume half-angle. Temperature may only affect the thermal velocity of the droplets, which is negligible compared to the drift velocity $ v_d $ of the droplets themselves. Therefore, changes in temperature may also not affect the plume half-angle.

\begin{figure}
    \centering 
    \includegraphics[width=0.8\linewidth]{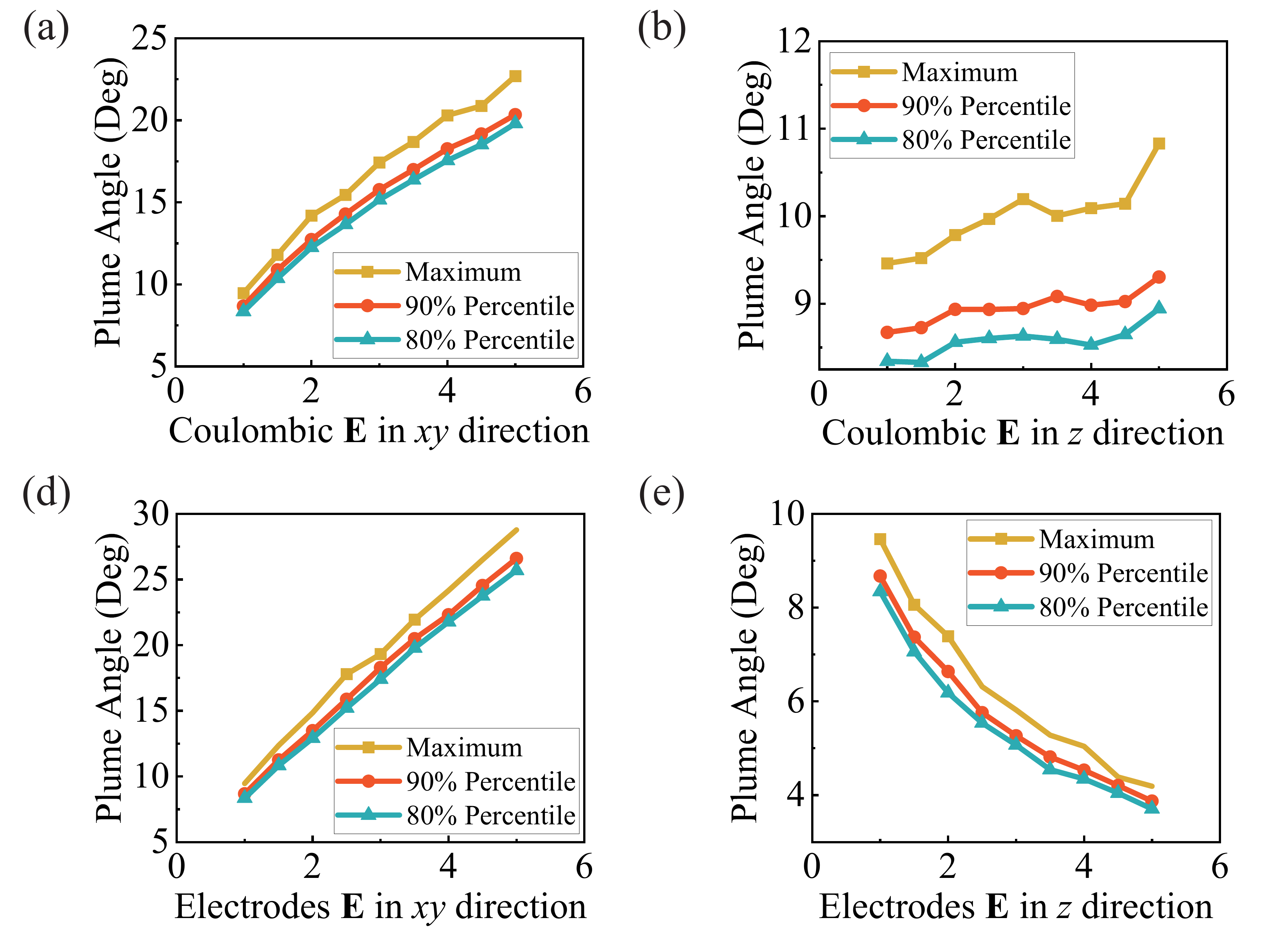}
    \caption{The impact of electric field on the half-angle of the plume.}
    \label{fig:plume angle - electric field}
\end{figure}

As shown in Fig.\ref{fig:plume angle - electric field}, to investigate the impact of different electric fields in various directions on the electrospray plume, only the electric field in one direction was modified and increased for all droplets during the simulation to amplify its effect. Fig.\ref{fig:plume angle - electric field}(a) explores the Coulomb electric field in the $xy$ direction; as the Coulomb electric field in the $xy$ direction increases, the half-angle of the plume noticeably increases, indicating that the radial Coulomb interaction is the main factor in the expansion of the electrospray plume. Fig.\ref{fig:plume angle - electric field}(b) explores the Coulomb electric field in the $z$ direction; as the Coulomb electric field in the $z$ direction increases, the half-angle of the plume slightly increases, but the increase is not significant, suggesting that the Coulomb electric field in the $z$ direction primarily hinders the downstream movement of droplets rather than accelerating their movement downstream. Fig.\ref{fig:plume angle - electric field}(c) explores the background electric field in the $xy$ direction; as the background electric field in the $xy$ direction increases, the half-angle of the plume noticeably increases. Fig.\ref{fig:plume angle - electric field}(d) explores the background electric field in the $z$ direction; as the background electric field in the $z$ direction increases, the half-angle of the plume noticeably decreases.

In summary, reducing the droplet charge, increasing the droplet mass, increasing the droplet emission time interval, and increasing the initial drift velocity of the droplets will result in a smaller half-angle of the electrospray plume. Within the plume, the radial Coulomb electric field and background electric field are the main factors contributing to plume expansion, while the axial Coulomb electric field hinders the downstream movement of droplets, and the axial background electric field accelerates droplet movement, thereby suppressing plume expansion.

\subsection{Fitting analysis of thruster parameters}

To further quantify the impacts of droplet charge, droplet mass, droplet emission time interval, and initial droplet velocity on the half-angle of the electrospray plume, a linear fitting method was employed to analyze the simulation results. Linear fitting, a commonly used statistical technique, determines the linear relationship between two variables by fitting data points. Specifically, the method of least squares was used to obtain the linear relationships between each factor and the plume half-angle.

The basic form of linear fitting is $y=ax+b$, where $y$ represents the plume half-angle, $x$ represents the influencing factor (such as droplet charge or mass), $a$ is the slope indicating the degree of influence of the factor on the plume half-angle, and $b$ is the $y$-intercept. The slope and intercept obtained from the fitting quantitatively describe the impact of each factor on the plume half-angle.

\begin{figure}[ht]
  \centering
  \includegraphics[width=0.8\linewidth]{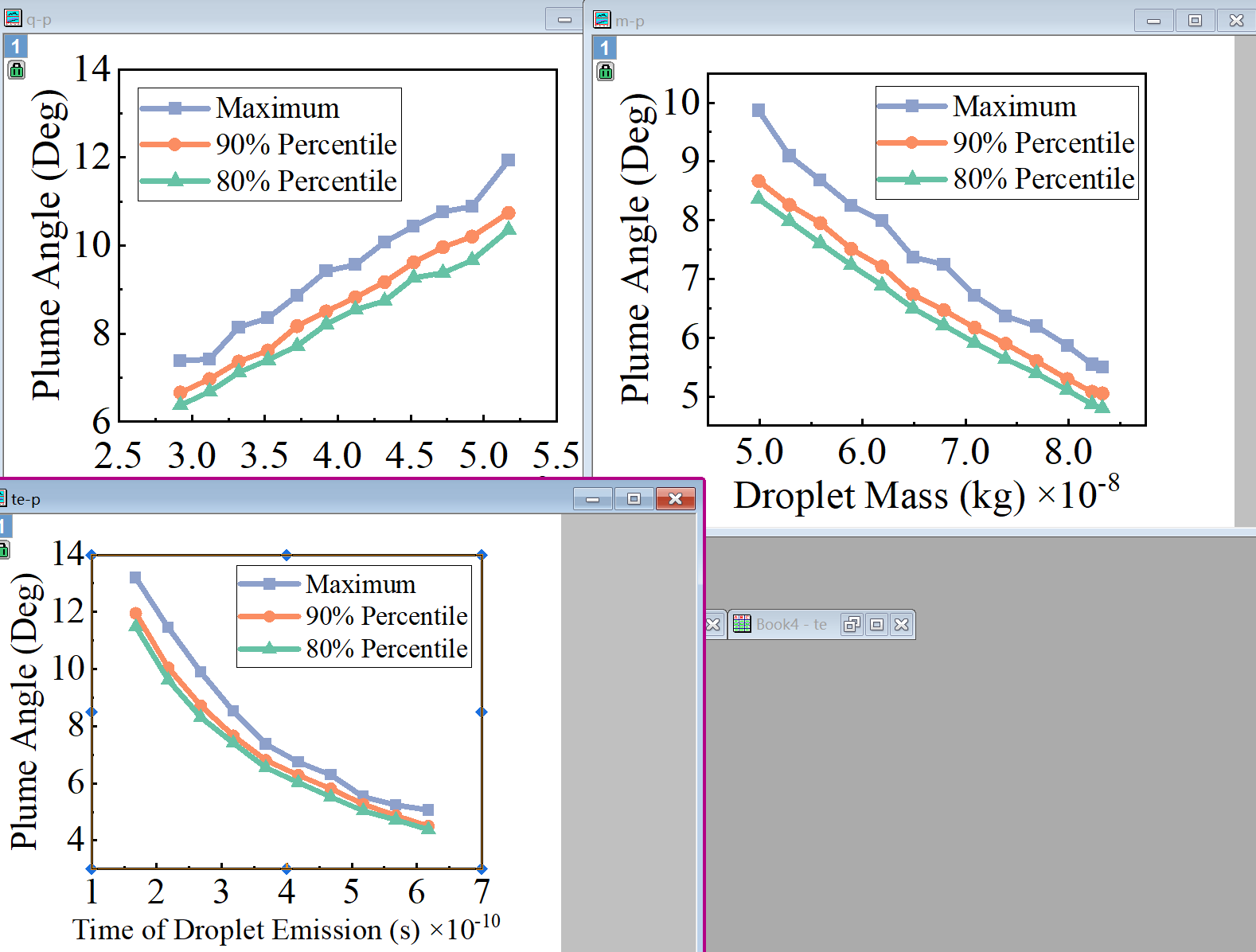}
  \caption{The quantitative correlations between plume half-angle and four key parameters - (a) droplet charge (linear fitting), (b) droplet mass (linear fitting), (c) droplet emission time interval (linear fitting), and (d) initial droplet velocity (nonlinear fitting) - are systematically characterized through curve fitting analysis.}
  \label{fig:linear fitting}
\end{figure}

\begin{table}[htbp]
	\centering
	\begin{tabular}{cccccc}
 \hline
Parameter & Function & $a$& $b$ & $R^2$ & $P$\\
\hline
 $q$ &$y=a+bx$& 1.30035 & 1.82619 & 0.99818 & $<0.0001$ \\
 $m$ &$y=a+bx$& 13.94927 & -1.08551  & 0.99439 &$<0.0001$ \\
 $\Delta t$ &$y=ax^b$& 8.98277 & -0.72928 & 0.99816 &$<0.0001$ \\
 $v_d$ &$y=a+bx$& 19.43824 & -0.01820 & 0.99011 &$<0.0001$ \\
  \hline
		\multicolumn{1}{l}{} & \multicolumn{1}{l}{} & \multicolumn{1}{l}{} & \multicolumn{1}{l}{}
	\end{tabular}
	\caption{Fitting parameters (\textit{a}, \textit{b}) and fitting indicators ($R^2$, $P$) for droplet-plume half-angle correlations}
	\label{tab:linear fitting}
\end{table}

In Fig.\ref{fig:linear fitting} and Tab.\ref{fig:linear fitting}, the fitting results are presented. Fig.\ref{fig:linear fitting} illustrates the linear and nonlinear correlations between plume half-angle and four parameters: droplet charge, droplet mass, droplet emission time interval, and initial droplet velocity. Tab.\ref{tab:linear fitting} summarizes the fitting coefficients ($a$, $b$) and correlation coefficients ( $R^2$ ) for each parameter. Analysis reveals that droplet charge, droplet mass, and initial droplet velocity exhibit linear relationships  with plume half-angle($y=a+bx$), whereas the droplet emission time interval follows a nonlinear correlation modeled with a power function ($y=ax^b$). All four fitting functions demonstrate exceptional accuracy, with $R^2 >0.99$ and $P < 0.0001$.

The positive linear correlation between droplet charge and plume half-angle indicates enhanced inter-droplet Coulomb interactions induce more pronounced plume expansion. Conversely, both droplet mass and initial velocity exhibit negative linear correlations with plume half-angle. Increased droplet mass reduces the electric field-induced acceleration, thereby suppressing plume divergence. Similarly, higher initial velocities limit the acceleration duration within the electric field, resulting in constrained plume expansion. The nonlinear inverse relationship between emission time interval and plume half-angle is hypothesized to result from reduced droplet population density per unit time, which weakens Coulomb interactions and subsequently inhibits plume expansion.

\section{Conclusions}

For the electrospray thruster, this paper uses a computationally precise particle-particle simulation method to simulate the electrospray plume and analyze the velocity of the plume droplets. The simulated electrospray thruster has a thrust of approximately $1.97 \times 10^{-6}$ N and a specific impulse of about 101 s. Subsequently, by controlling variables and changing one parameter at a time, the influence of various parameters on the expansion of the electrospray plume is investigated using the plume droplet half-angle. The conclusion is that reducing the droplet charge, increasing the droplet mass, increasing the droplet emission time interval, and increasing the initial drift velocity of the droplets will result in a smaller electrospray plume half-angle. Both the radial Coulomb electric field and the background electric field are the main factors in plume expansion. 
Fitting analysis between various thruster operating parameters and electrospray plume angles reveals three distinct relationships: (1) droplet charge shows a positive linear relationship with plume angle ($y=a+bx,\  b>0$), (2) droplet mass and initial drifting velocity both exhibit negative linear relationships with plume angle ($y=a+bx,\  b<0$), while (3) droplet emission time interval demonstrates an exponential decay relationship with plume divergence ($y=ax^b$). These quantitative dependencies offer critical guidance for thruster design optimization targeting plume control.

\section*{Acknowledgments}

The authors acknowledge the support from National Natural Science Foundation of China (Grant No.5247120164).

\section*{Data availability}

The data that support the findings of this study is available from the corresponding author upon reasonable request.

\section*{Reference}

\bibliographystyle{elsarticle-num}
\bibliography{reference}

\end{document}